\documentclass[%
 aip,
 apl,%
 amsmath,amssymb,
 reprint,%
]{revtex4-1}

\usepackage{graphicx}
\usepackage{dcolumn}
\usepackage{bm}
\usepackage{booktabs}
\usepackage{upgreek}
\usepackage[colorlinks=true,linkcolor=blue,citecolor=blue,allcolors=blue]{hyperref}
\usepackage{color,soul}

\begin{document}
\title{Control of Magnetic Order in Spinel ZnFe$_2$O$_4$ Thin Films Through Intrinsic Defect Manipulation}

\author{V. Zviagin}
\email{vitaly.zviagin@uni-leipzig.de}

\affiliation{ Universit\"{a}t Leipzig, Felix-Bloch-Institut f\"{u}r Festk\"{o}rperphysik, Linn\'{e}stra\ss e 5, D-04103 Leipzig, Germany}

\author{C. Sturm}

\affiliation{ Universit\"{a}t Leipzig, Felix-Bloch-Institut f\"{u}r Festk\"{o}rperphysik, Linn\'{e}stra\ss e 5, D-04103 Leipzig, Germany}

\author{P. Esquinazi}

\affiliation{ Universit\"{a}t Leipzig, Felix-Bloch-Institut f\"{u}r Festk\"{o}rperphysik, Linn\'{e}stra\ss e 5, D-04103 Leipzig, Germany}

\author{M. Grundmann}

\affiliation{ Universit\"{a}t Leipzig, Felix-Bloch-Institut f\"{u}r Festk\"{o}rperphysik, Linn\'{e}stra\ss e 5, D-04103 Leipzig, Germany}

\author{R. Schmidt-Grund}

\affiliation{ Universit\"{a}t Leipzig, Felix-Bloch-Institut f\"{u}r Festk\"{o}rperphysik, Linn\'{e}stra\ss e 5, D-04103 Leipzig, Germany}

\affiliation{Technische Universit\"{a}t Ilmenau, Institut f\"{u}r Physik, Weimarer Stra\ss e 32, D-98684 Ilmenau, Germany}

\begin{abstract}

We present a systematic study of the magnetic properties of semiconducting ZnFe$_2$O$_4$ thin films fabricated by pulsed laser deposition at low and high oxygen partial pressure and annealed in oxygen and argon atmosphere, respectively. The magnetic response is enhanced by annealing the films at 250{$^{\circ}$C} and diminished at annealing temperatures above 300{$^{\circ}$C}. The initial increase is attributed to the formation of oxygen vacancies after argon treatment, evident by the increase in the low energy absorption at $\sim$ 0.9\,eV involving Fe$^{2+}$ cations. The weakened magnetic response is related to a decline in disorder with a cation redistribution toward a normal spinel configuration. The structural renormalization is consistent with the decrease and increase in oscillator strength of respective electronic transitions involving tetrahedrally (at $\sim$ 3.5\,eV) and octahedrally (at $\sim$ 5.7\,eV) coordinated Fe$^{3+}$ cations.

\end{abstract}

\date{\today}

\maketitle


Spinel ferrite, ZnFe$_2$O$_4$, is identified as a paramagnetic insulator with an antiferromagnetic order below the N{\'{e}}el temperature of $\sim$10\,K.\cite{1996} It crystallizes in a normal spinel configuration, where Zn$^{2+}$ and Fe$^{3+}$ cations are distributed over tetrahedral (\textit{Td}) and octahedral (\textit{Oh}) coordinated intersitial sites of the anion \textit{fcc} lattice, respectively. While the sole magnetic interaction in this state is expected to be the antiferromagnetic (AF) oxygen mediated superexchange (SE) interaction between Fe$^{3+}_{Oh}$ cations, single crystal ferrite was shown to exhibit strong geometrical frustration.\cite{Kamazawa2003} The formation of local defects is likely to relieve the frustration and give rise to electrical conductivity and absorption in the visible spectral range due to the presence of Fe$^{2+}$, as well as spontaneous magnetization at room temperature.\cite{Brachwitz2013,Marcu2007,Bontgen2013,RodriguezTorres2011,Stewart2007} Therefore, the semiconducting and magnetic properties can be tailored by manipulating the type and concentration of intrinsic defects, which make spinel ferrite particularly attractive for spintronic device engineering, energy storage, optoelectronic and biomedical applications.\cite{Taffa2012,Hirohata2014,Abdi2017,Varzi2014,Yang2017,Guijarro2017,Lee2013,Meidanchi2015,Sawant2016,Venkateshvaran2009}

The ferrimagnetic order in ZnFe$_2$O$_4$ thin films is commonly attributed to the presence of tetrahedrally coordinated Fe$^{3+}$ cations as a result of the inversion mechanism or due to Fe$^{3+}$ on nominally unoccupied tetrahedral lattice sites.\cite{Stewart2007,RodriguezTorres2014} This gives rise to the dominant AF oxygen mediated SE interaction between Fe$^{3+}$ cations of tetrahedral and octahedral coordination (\textit{Oh}-\textit{Td}).\cite{RodriguezTorres2014} The increase of the magnetic order with the decrease in fabrication temperature was previously correlated to the enhanced amplitude of the electronic transition, involving Fe$^{3+}_{Td}$ cations.\cite{Zviagin2016a} However, precise assignment of visible transitions remains controversial.\cite{Schlegel1979,Shinagawa1992,Zviagin2016b,Moskvin2010,Ziaei2017,Fritsch2018} Furthermore, oxygen deficient spinel ferrites displayed a strong magnetic response when grown at low oxygen partial pressure or annealed in vacuum.\cite{RodriguezTorres2011,Jedrecy2014,Tanaka2006,Nakashima2007,Sultan2009} In this case, the oxygen vacancy formation would cause an imbalance of the O-Zn and O-Fe bond strength, thus leading to the distortion of the unit cell structure.\cite{Yao2006} As a result, the oxygen mediated AF coupling (\textit{Oh}-\textit{Oh}) would then be modified to create a strong ferromagnetic (FM) interaction mediated by an oxygen vacancy.\cite{RodriguezTorres2014} Due to the limited spectroscopic evidence and understanding of the interplay between the individual defects, the magnitude of their contribution to the spin configuration and magnetic behavior remains unclear. 

In this work, we present the magnetic properties of disordered ZnFe$_2$O$_4$ thin films in dependence on the oxygen partial pressure during fabrication as well as thermal annealing temperature and atmosphere. Based on the strength of the electronic transitions, visible in the parametric model dielectric function, we demonstrate that the magnetic response can be enhanced by the formation of oxygen vacancies and reduced by cation redistribution toward a normal spinel configuration.


The investigated thin films were fabricated by pulsed laser deposition (PLD), similar to the procedure described in Ref. \onlinecite{Lorenz2011}. The (100) SrTiO$_3$ substrate temperature was kept constant at $\sim$300{$^{\circ}$}\,{C} and $\sim$70\,nm thick ZnFe$_2$O$_4$ (ZFO) films were grown at low (6 $\times$ 10$^{-5}$\,{mbar}) and high (0.016\,{mbar}) oxygen partial pressure, denoted by LP ZFO and HP ZFO, respectively. The LP ZFO and HP ZFO samples were divided into four pieces and annealed in the temperature range from 250 to 375\,{$^\circ$C} in 100\,mbar oxygen (O250-O375) and argon (Ar250-Ar375) atmosphere, respectively. A low (high) deposition pressure and an oxygen (argon) annealing atmosphere was chosen with the intention to eliminate (facilitate) oxygen vacancies in our thin films. In order to check the reproducibility induced by thermal annealing, magnetization of two nominally identical (Ar250) films was measured and a deviation within 4.7\% was determined. The annealing time of 4 hours was chosen, after which the magnetic response saturates, as determined previously.\cite{Kumar2017} 

The crystalline structure was analyzed by X-ray diffraction (XRD) $2\theta - \omega$ scans using a wide-angle Phillips X'Pert Bragg-Brentano diffractometer with Cu K$_{\alpha}$ radiation. Field dependent magnetic properties of the thin films were recorded using a physical property measurement system (PPMS-Quantum Design, USA) operating in vibrating sample magnetometer (VSM) mode. Magnetization as a function of temperature was measured under field-cooled (FC) and zero-field-cooled (ZFC) procedure at an external field of 0.1\,{T}, applied in the sample surface plane. 

The thin film complex dielectric function (DF) was obtained using a spectroscopic ellipsometer in a polarizer-compensator-sample-analyzer (PCSA) configuration. Measurements were carried out at ambient conditions in the spectral range from 0.5 to 8.5\,eV and at angles of incidence of 60\,$^{\circ}$ and 70\,$^{\circ}$. The spectra were analyzed by means of transfer matrix technique for a layer stack model consisting of layers for the substrate, the spinel material, and the surface. The DF of the substrate was determined prior to film deposition. The DF of the surface layer was described by Bruggeman effective medium approximation, consisting of a mixture of the thin film DF and that of void and correlated to surface morphology measurements, depicted in Fig. S1.\cite{Lehmann2014} The film thickness and the DF line-shape was initially determined using a numerical B-Spline approximation.\cite{LIKHACHEV2017519} A parametric model DF (MDF), composed of a series of Gaussian, Lorentzian and M0-critical point model function approximations, was derived and regression analysis was applied for the best match to the numerical DF as well as the experimental data (see supplementary material for further details).


\begin{figure}[htb!]
\begin{center}
\includegraphics[width=1\columnwidth]{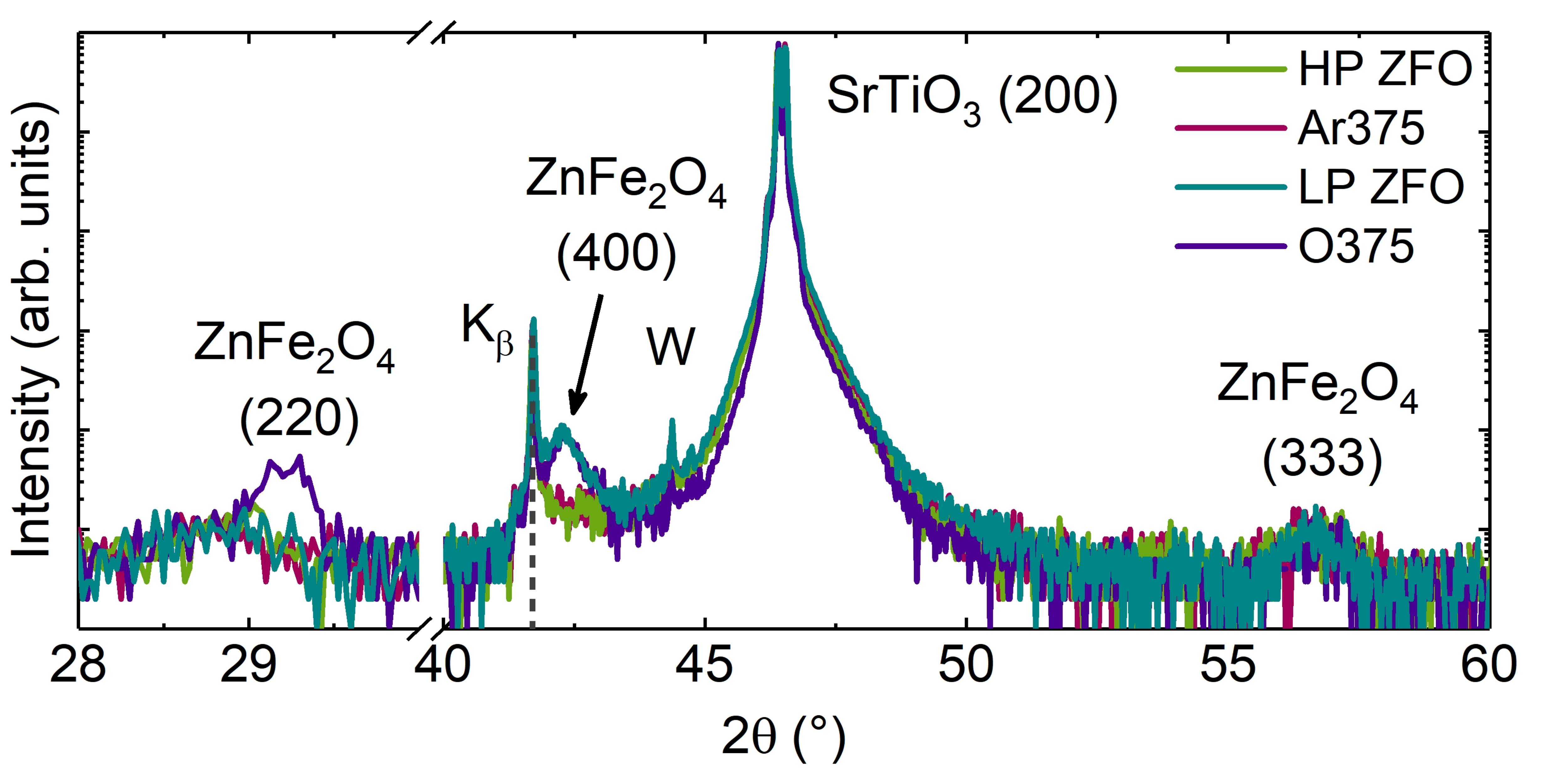} \caption{ XRD wide-angle $2\theta$ scans for as-deposited films and after annealing at 375{$^{\circ}$}\,{C}. Substrate reflexes, marked by K$_{\beta}$ and W, correspond to Cu K$_{\beta}$ and W L$_{\alpha}$ spectral lines of the X-ray tube, respectively.}
\label{Fig1}
\end{center}
\end{figure}

XRD $2\theta - \omega$ scans for the as-deposited films and after annealing at 375{$^{\circ}$}\,{C} are depicted in Fig. \ref{Fig1}. All films demonstrate a weak (333) ZnFe$_2$O$_4$ reflex. The HP ZFO film shows no additional peaks and no changes upon thermal treatment. The LP ZFO film, on the other hand, exhibits a (400) reflex, suggesting a (100) orientated ZnFe$_2$O$_4$ growth on the (100) SrTiO$_3$ substrate. After annealing LP ZFO at the highest temperature, the O375 film shows a secondary phase of (110) orientation, while the (400) reflex remains unperturbed by thermal treatment.

\begin{figure}[htb!]
\begin{center}
\includegraphics[width=1\columnwidth]{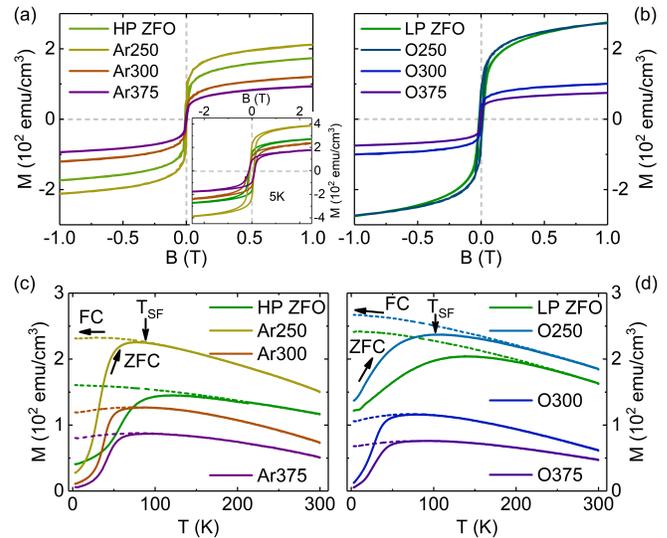} \caption{Volume magnetization measured at 300\,{K} as a function of applied magnetic field, B, for HP and LP ZFO films annealed in (a) argon and (b) oxygen atmosphere, respectively. The inset in (a) shows measurements at 5\,K for the HP ZFO film annealed in argon. Volume magnetization as a function of temperature, at external field of B = 0.1\,T, for HP and LP ZFO films annealed in (c) argon and (d) oxygen atmosphere, respectively. The arrows correspond to the FC or ZFC measurement procedure and the spin freezing temperature T$_{SF}$ is exemplary indicated for one film in each series. }
\label{Fig2}
\end{center}
\end{figure}

Fig. \ref{Fig2} (a) and (b) show volume magnetization as a function of applied magnetic field measured at room temperature as well as 5\,K (inset in Fig. \ref{Fig2} (a)). Room temperature magnetization saturation as well as the "S" shape of the magnetic hysteresis is observed for as-deposited films and after annealing at 250\,{$^\circ$C}. While the magnetic response is enhanced after annealing HP ZFO at 250\,{$^\circ$C}, the LP ZFO exhibits a more pronounced hysteresis shape, suggesting a magnetically harder material after oxygen treatment at 250\,{$^\circ$C}, Fig. \ref{Fig2} (b). Both films exhibit a weakened magnetic response with the increase in annealing temperature $\geq$300\,{$^\circ$C}. In comparison to the measurements at 300\,K, the magnetization measurements at 5\,K display a similar trend upon thermal treatment along with a notably pronounced coercive field. The coercive field, listed in Table \ref{Table_Ann}, ranges from 82.8 (Ar250) to 132.1\,{mT} (O375), with a general increase with increase in thermal treatment temperature. It is important to note that after subtraction of the diamagnetic contribution, arising from the substrate, a saturation of magnetization was not achieved for films annealed at temperatures $\geq$300\,{$^\circ$C} in either atmosphere. The linear magnetization behavior at high fields could suggest the presence of a paramagnetic component, as recently reported for a similar system.\cite{Zamiri2017,SalcedoRodriguez2018}

Volume magnetization was recorded as a function of temperature in the FC and ZFC mode with an applied field of B = 0.1\,T, Fig. \ref{Fig2} (c,d). First and foremost, the ZFC curve for all samples demonstrates a maximum magnetization value at a temperature below room temperature. Since ZnFe$_2$O$_4$ is classified as a geometrically frustrated compound, which shows spin- or cluster-glass magnetic behavior, this temperature is henceforth referred to as the spin freezing temperature (T$_{SF}$).\cite{SalcedoRodriguez2018} Furthermore, magnetic irreversibility below T$_{SF}$ is visible for all samples and the difference between the ZFC and FC curves decreases with the increase in thermal treatment temperature. Lastly, a convex magnetization behavior, a cusp, apparent in the ZFC curves below T$_{SF}$, becomes more pronounced upon annealing in argon rather than in oxygen atmosphere. The described magnetization behavior of films annealed at temperatures $\geq$330\,{$^\circ$C} is characteristic of a near-to-bulk normal spinel structure with antiferromagnetic spin alignment, therefore a lack of magnetic moment is expected at temperatures below the N{\'{e}}el temperature ($\leq$12\,K).\cite{Kamazawa2003,Nakashima2007,MeloQuintero2019}

\begin{figure}[htb!]
\begin{center}
\includegraphics[width=1\columnwidth]{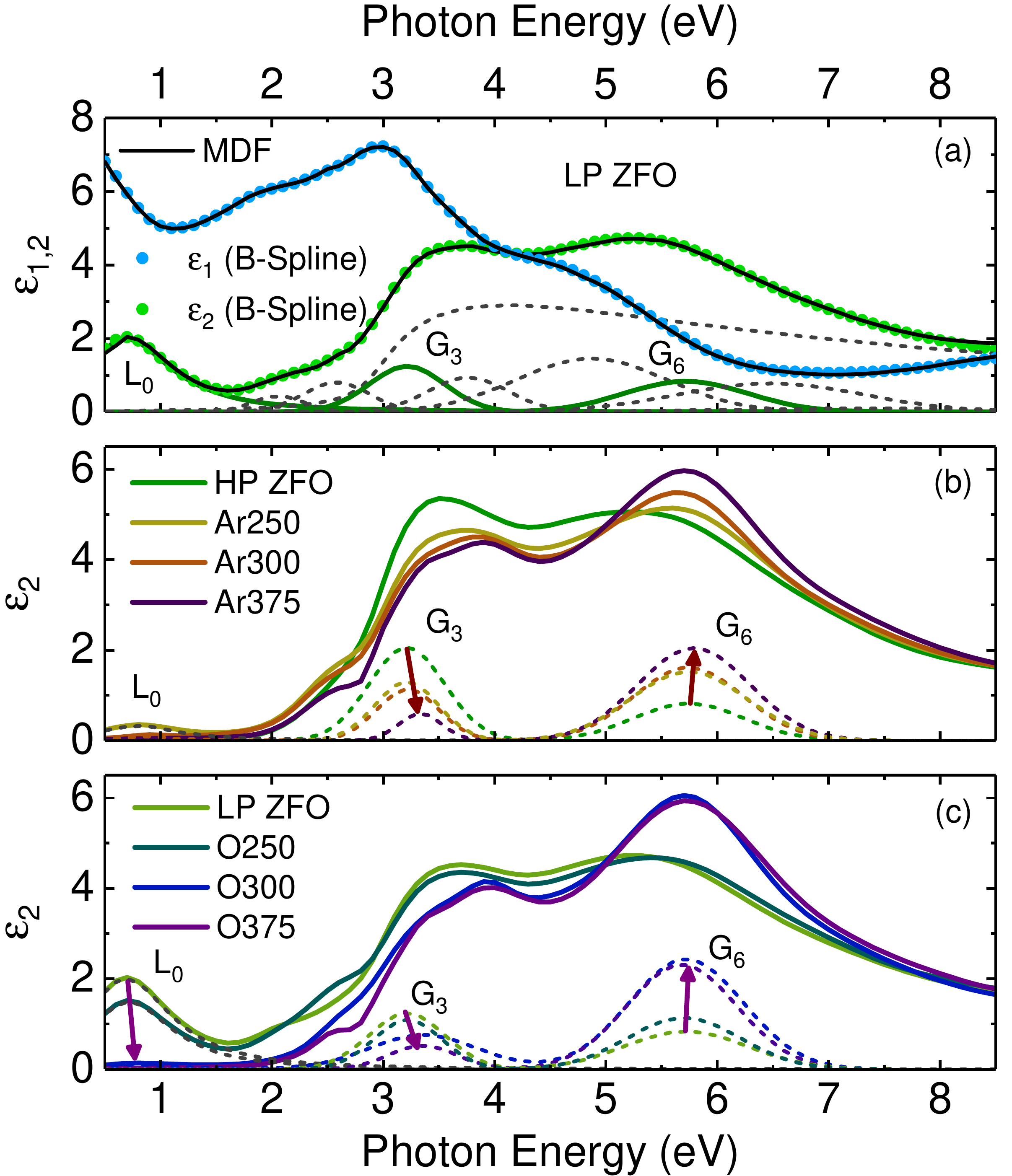} \caption{(a) The numerical (B-Spline) and parametric (MDF) approximation of the complex DF line-shape for the LP ZFO film, depicted by points and solid lines, respectively. Individual function contributions to the MDF line-shape are exemplary depicted for the imaginary component. MDF of films annealed in (b) argon and (c) oxygen atmosphere. Contributions of the L$_0$, G$_3$ and G$_6$ approximation functions to the MDF are depicted by solid lines in (a) and dashed lines in (b) and (c), see text for more details. }
\label{Ann_MDF}
\end{center}
\end{figure}

The bulk cationic distribution is probed by spectroscopic ellipsometry and investigated based on the strength of electronic transitions visible in the complex DF, $\tilde\varepsilon = \varepsilon_{1} + {\emph {i}}\varepsilon_{2}$.\cite{Zviagin2016a,Zviagin2016b} The fit of the parametric model dielectric function (MDF) to the numerical approximation (B-Spline) for LP ZFO as well as individual function contributions to the MDF ($\varepsilon_2$) are depicted in Fig. \ref{Ann_MDF} (a). A significant absorption in the low energy range is exhibited by the LP ZFO thin film, denoted by L$_0$ in Fig. \ref{Ann_MDF} (a). Based on the energy of the absorption feature ($\sim$0.9\,eV), it is related to an electronic transition between \textit{d} orbitals of Fe$^{2+}_{Oh}$ cations.\cite{PhysRevB.56.5432,Schlegel1979} A decrease in the amplitude of this transition is observed upon annealing the LP ZFO film in oxygen, Fig. \ref{Ann_MDF} (c). Interestingly, an increase in the low energy absorption is evident in the MDF of HP ZFO after annealing in argon at 250\,{$^\circ$C}, indicating a notable concentration of Fe$^{2+}_{Oh}$ cations likely present in Ar250 thin film.

The absorption feature approximated by a Gaussian oscillator and located at $\sim$3.5\,eV has been previously shown to involve tetrahedrally coordinated Fe$^{3+}$ cations.\cite{Zviagin2016b,Ziaei2017} It exhibits a strong magneto-optical response and can be assigned to either a charge transfer transition between Fe$^{3+}_{Oh}$ and Fe$^{3+}_{Td}$, or an electronic transition between O 2\textit{p} to 3\textit{d} orbital of Fe$^{3+}_{Td}$.\cite{Zviagin2016a,Liskova-Jakubisova2015,Ziaei2017} The decrease in amplitude with the increase in deposition temperature was directly correlated to the decrease in ferrimagnetic order by weakening the dominating AF \textit{Oh}-\textit{Td} interaction.\cite{Zviagin2016b} Similar behavior is observed in the present study. A decline in the amplitude of the function, denoted by G$_3$ in Fig. \ref{Ann_MDF} (b,c), is observed with the rise in treatment temperature. Simultaneously, the amplitude of the absorption feature, located at $\sim$5.7\,eV and denoted by G$_6$ in Fig. \ref{Ann_MDF} (b,c), is enhanced with the increase in annealing temperature. This function suggests a change in the concentration of Fe$^{3+}_{Oh}$ cations and can be assigned to an electronic transition between O 2\textit{p} and 4\textit{s} band of Fe$^{3+}_{Oh}$.\cite{Shinagawa1992,Zviagin2016b,Moskvin2010,Fritsch2018} The decrease (increase) in the amplitude of transition, involving Fe$^{3+}_{Td}$ (Fe$^{3+}_{Oh}$) cations, reveals a cationic redistribution towards a less inverted spinel structure with the increase in annealing temperature.

\begin{figure}[htb!]
\begin{center}
\includegraphics[width=1\columnwidth]{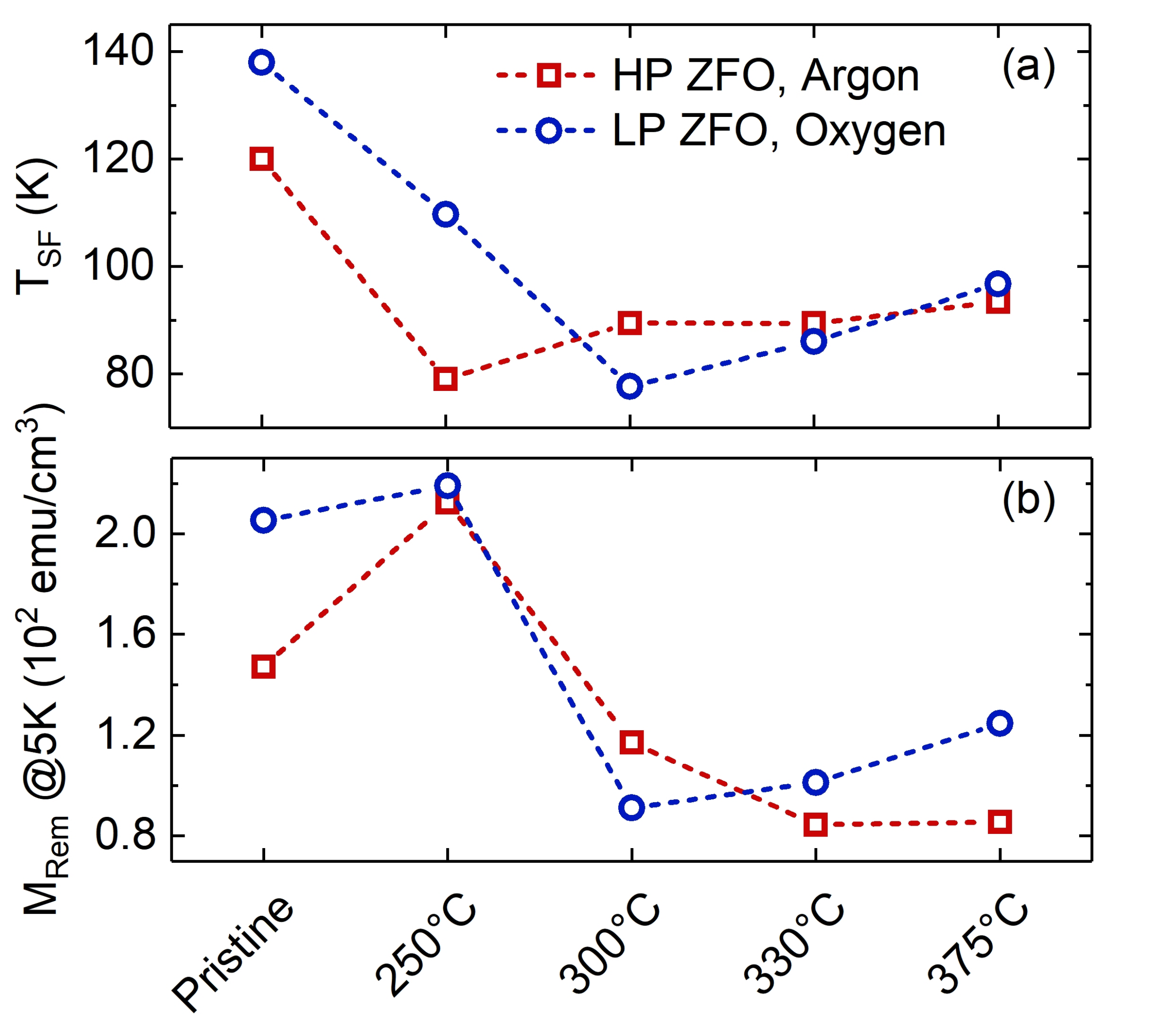} \caption{ (a) Spin freezing temperature T$_{SF}$ and (b) remanent magnetization M$_{Rem}$, measured at 5\,K, as a function of annealing temperature. The red and blue points correspond to HP ZFO and LP ZFO annealed in argon and oxygen atmosphere, respectively. }
\label{Ann_Diss}
\end{center}
\end{figure}

The spin freezing temperature, T$_{SF}$, as a function of annealing temperature is depicted in Fig. \ref{Ann_Diss} (a) and listed in Table \ref{Table_Ann}. A decrease from 120 to 79\,K in T$_{SF}$ is observed upon annealing HP ZFO in argon atmosphere at 250\,{$^\circ$C}. This behavior can be explained on the basis of a less disordered cationic structure in the Ar250 film.\cite{Nakashima2007,SalcedoRodriguez2018} The decline in G$_3$ and an enhanced G$_6$ function amplitude upon annealing HP ZFO in argon further validates this assumption and a decrease in magnetic response is anticipated. However, an increase in the remanent magnetization (5\,K) for Ar250 thin film is observed, Fig. \ref{Ann_Diss} (b). Simultaneously, absorption in the low energy spectral range becomes apparent and would suggest the presence of Fe$^{2+}$ cations. Therefore, the enhanced magnetization can be explained by the formation of oxygen vacancies, which would strengthen the FM \textit{Oh}-\textit{Oh} interaction and contribute to the overall ferrimagnetic order.\cite{RodriguezTorres2014}

Only a slight increase in the M$_{Rem}$ and a decrease in T$_{SF}$ is observed upon treating the LP ZFO film in oxygen at 250\,{$^\circ$C}. This could be explained by a decrease in Fe$^{2+}$ cation concentration due to the treatment in oxygen atmosphere. In this case, the oxygen anions would contribute to the decrease in cation disorder by producing Fe$_{Oh}^{3+}$ cations and strengthening the AF \textit{Oh}-\textit{Td} interaction. 

The lowest value of T$_{SF}$ as well as M$_{Rem}$ is exhibited by annealing the LP ZFO film in oxygen at 300\,{$^\circ$C}. This is attributed to the cation redistribution toward a less defective configuration, coincident with the decline and rise in the amplitude of G$_{3}$ and G$_{6}$ functions, respectively. Interestingly, an increase in T$_{SF}$ as well as M$_{Rem}$ is observed upon annealing in oxygen at temperatures $\geq$300\,{$^\circ$C}. A considerable increase in the magnetic response has been recently demonstrated for ZnFe$_2$O$_4$ annealed in vacuum at temperatures $\geq$450\,{$^\circ$C} and attributed to an oxygen deficient environment during the process of inversion recovery.\cite{SalcedoRodriguez2018} In our case, however, annealing HP ZFO in argon at temperatures $\geq$300\,{$^\circ$C} results in a nearly constant T$_{SF}$, while a decrease in M$_{Rem}$ is apparent. 

It is important to note, the formation of magnetic clusters in regions with different degrees of inversion can not be excluded and additional characterization methods are necessary. Nevertheless, consistent with the cations active in optical absorption, and in relation to the magnetic behavior, the following cation distribution can be assumed:\cite{Brachwitz2013}

\begin{equation}
{ (Zn^{2+}_{1-\lambda} Fe^{3+}_{\lambda})_{Td} (Zn^{2+}_{\lambda} Fe^{2+}_{\delta} Fe^{3+}_{2-\lambda-\delta})_{Oh} O_4^{2-} }
\label{eq1}
\end{equation}

\noindent where $\delta$ would correspond to the concentration of Fe$^{2+}_{Oh}$ cations, which are induced by oxygen vacancy formation and show a strong dependence on fabrication and treatment atmosphere. The parameter $\lambda$ ($0 \leq \lambda \leq 1$) is the inversion parameter and corresponds the inversion mechanism, where Zn$^{2+}_{Td}$ and Fe$^{3+}_{Oh}$ are displaced.\cite{RodriguezTorres2014} The deviation from the inversion parameter is possibly due to Fe$^{3+}$ occupying nominally unoccupied \textit{Td} lattice sites. The concentration of this defect declines with increasing treatment temperature, independent of the annealing atmosphere.

In summary, we demonstrate that the magnetic properties of ZnFe$_2$O$_4$ thin films comprising of defects can be manipulated based on the oxygen partial pressure during fabrication as well as thermal treatment temperature and atmosphere. The formation of oxygen vacancies was found to be responsible for the enhanced magnetic response in Ar250, as evident by the low energy absorption $\sim$0.9\,eV involving Fe$^{2+}$ cations. Diminished magnetic response is likely due to cation redistribution ($\geq$300\,{$^\circ$C}), consistent with a decrease (increase) in amplitude of the electronic transition involving tetrahedrally (octahedrally) coordinated Fe$^{3+}$ cations. Based on the cations involved in optical absorption, our results reveal a deeper understanding of the fundamental interactions responsible for the defect induced magnetism in semiconducting spinel ferrite compound.





\begin{table}[htb!]
\caption{\label{Table_Ann} Values of the coercivity field (H$_{Coer}$), remanent magnetization (M$_{Rem}$), both measured at 5\,K as well as spin freezing temperature (T$_{SF}$) for the corresponding thin films.  }
\begin{center}
\begin{tabular}{cccc}

\hline
\hline
Sample &  H$_{Coer}$ (\,mT) & T$_{SF}$ (\,K) & M$_{Rem}$ (\,emu/cm$^3$) \\

\hline

HP ZFO & 96.5  & 120.0 & 147.2  \\

Ar250  & 82.8 & 79.1 & 212.3  \\

Ar300  &  125.8 & 89.5 & 117.1 \\

Ar330  & 119.1 & 89.4 & 84.5  \\

Ar375 & 117.5 & 93.4 & 85.5 \\

\hline

LP ZFO & 103.8 & 138.0 & 205.3 \\

O250  & 95.5 & 109.7 & 219.1  \\

O300  & 104.9 & 77.7 & 91.2 \\

O330  & 107.4 & 86.1 & 101.2  \\

O375  & 132.1 & 96.8 & 124.7 \\

\hline

\end{tabular}
\end{center}
\end{table}

See supplementary material for the thin film surface morphology, model fit to the measured ellipsometric parameters as well as the table of resonance energies of the MDF contributions and respectively assigned transitions.

We gratefully acknowledge Gabriele Ramm for PLD target preparation and Holger Hochmuth for the thin film growth. This work was funded by the Deutsche Forschungsgemeinschaft (DFG, German Research Foundation) – Projektnummer 31047526 – SFB 762, projects B1, B3, A7. The authors declare no conflict of interest.

\bibliography{ZFO_Annealed_APL}


\end{document}